# Language Non-Invariance in the Ideological Evaluation of Large Language Models

Eduardo Levy Yeyati* César Ciappa† Milagros Onofri‡

**Abstract**

Recent multilingual studies show that political evaluations of large language models (LLMs) can vary across languages. This paper examines how robust this non-invariance is to questionnaire translation, surface wording, stochastic model variation, and instrument choice, and how much it matters for cross-model comparison. Using the Political Compass, the analysis compares the positions of five LLMs, GPT-3.5, GPT-4o, GPT-5.2, Qwen-2.5, and Mistral Large, on its economic and social axes in English, Spanish, and French, with repeated stochastic draws. Additional exercises use independent Spanish and French translations, paraphrased versions in all three languages, and an alternative ideological instrument. Measured ideological positions are not language-invariant: relative to English, between four and five of the ten pairwise model rankings reverse in Spanish or French, depending on the language and axis. The broad cross-language pattern persists under independent translations, indicating that it is not solely an artifact of the official questionnaire versions. Yet the magnitude of particular shifts, and in some cases their sign, remains sensitive to wording and instrument choice. The evidence therefore supports a measurement interpretation: ideological audits recover language- and instrument-conditioned outcomes rather than a language-free political attribute. Because a single provider serves distinct linguistic constituencies through a common product, without users observing what other constituencies receive, this lack of comparability also bears on political accountability and cross-model comparisons.

*Department of Economics, Universidad Torcuato Di Tella, Buenos Aires, Argentina. E-mail: ely@utdt.edu.

†Corresponding author. Department of Economics, Universidad de San Andrés, Victoria, Buenos Aires, Argentina. E-mail: cciappa@udesa.edu.ar.

‡CEDLAS, Universidad Nacional de La Plata, La Plata, Buenos Aires, Argentina. E-mail: milagros.onofri@econo.unlp.edu.ar.

## 1 Introduction

Large language models increasingly summarize policy debates, answer politically charged questions, and mediate access to information for users in multiple languages. Their growing role has generated a literature on the political bias of LLMs (Motoki et al. 2024; Feng et al. 2023; Santurkar et al. 2023) and on differences across model generations (Rozado 2024). Much of this early work, however, evaluates models in English and implicitly treats language as a neutral channel through which stable ideological properties are revealed.

Treating language as a neutral channel is consequential because multilingual models are trained on data that differ across languages in volume, framing conventions, institutional vocabulary, and supervision after training. If the same ideological instrument yields different measured positions depending on the language of elicitation, part of what is interpreted as ideological change across model generations may instead reflect measurement that depends on language.

This issue also has a substantive dimension as large language models become political information intermediaries. A single provider serves users who speak English, Spanish, and French through the same branded system, while language partitions those users into constituencies that cannot observe one another. The political curation received by one group is not visible to the others, and users cannot compare the answers they receive with those delivered in a language they do not use. The collective choice dimension of this setting arises, in part, because alignment, localization, and political auditing are costly and their returns are unlikely to be symmetric across language markets. Regulatory scrutiny of political content, public benchmarks, and the academic audit literature operate almost exclusively in English; even audits that draw conclusions about electorates that do not primarily use English, such as the inferences for Brazil and the United Kingdom in Motoki et al. (2024), rely on prompts in English. A provider equating the marginal cost and marginal return of political curation across languages will therefore have stronger incentives to curate where scrutiny binds. At the same time, an asymmetry that cannot be observed by those who bear it cannot be disciplined through either exit or voice (Hirschman 1970): users cannot switch away from, or protest, a difference they cannot see. Comparability of ideological measurements across languages is thus not merely a refinement of audit methodology, but the monitoring technology on which the political accountability of a multilingual intermediary depends. We do not observe the alignment investments or objective functions of providers, so the design does not identify this mechanism on the supply side. What the paper provides is the comparability evidence that such accountability requires, together with descriptive patterns across providers and model generations, discussed in Section 6, that may reflect both supply-side incentives and demand from different linguistic groups, which our analysis cannot distinguish between.

Recent studies already show that political evaluations of LLMs can vary across languages. Helwe et al. (2025) administer machine translated versions of the Political Compass Test to fifteen models in 33 languages and find that changes in prompting language move measured placement more than an assigned nationality does. Using the official translations of the test and eleven paraphrased prompt templates around each statement, Gurgurov et al. (2025) reach a similar conclusion for seven open models in fourteen languages. Language dependent stances on the same 50 country benchmark are also reported by Nadeem et al. (2026) for a further set of proprietary and open models, while Rettenberger et al. (2025) document differences between English and German on a voting advice instrument. This pattern is not universal, however: using statements from voting advice applications in five Western European languages, Weeber et al. (2026) find very few significant differences across languages and conclude that political opinions largely transfer across languages. The lack of invariance across languages in LLM ideological measurement is therefore already established, but its extent is design dependent. The contribution here is not to claim priority for that finding, but to ask how much of the measured variation across languages survives changes in questionnaire translation, surface wording, and the ideological instrument, and how large it is relative to the stochastic variability of model responses.

The empirical design uses the full Political Compass questionnaire under a common protocol for five large language models, GPT-3.5, GPT-4o, GPT-5.2, Qwen-2.5, and Mistral Large, evaluated in English, Spanish, and French. In the main specification, models receive only the questionnaire instructions, without being assigned any political identity. Each questionnaire item is submitted in an independent call, with the same item order across languages, a temperature of 0.5, and four repetitions for every combination of model and language. The analysis is then extended using independent machine translations of the English questionnaire into Spanish and French, paraphrased versions of the questionnaire in all three languages, and an alternative ideological instrument.

The results show that measured ideological positions are not invariant across languages, and that the differences are large enough to affect model comparisons. Relative to English, between four and five of the ten pairwise model rankings reverse in Spanish or French, depending on the language and axis, although some of these reversals occur between models with closely estimated positions. The broad pattern also persists under independently produced Spanish and French translations, indicating that it is not solely an artifact of the official questionnaire versions, although individual estimates remain sensitive to the version used. Alternative wording and a different ideological instrument lead to the same general conclusion: differences across languages remain, while the magnitude of particular shifts, and in some cases their sign, changes with the measurement procedure.

Accordingly, the paper does not interpret these differences as evidence that models possess stable human political preferences that change with language. The estimates concern measured ideological placement, not language specific latent ideology. Their relevance lies in comparability: results obtained in one language and with one instrument cannot be assumed to carry over mechanically to another. Methodologically, the paper combines matched comparisons at the item level with stochastic repeated draws, allowing variation across runs to be assessed alongside sensitivity to questionnaire composition, and then examines whether the results survive changes in translation, wording, and instrument. It also quantifies the implications for comparisons across models. The contribution is therefore evaluative rather than an attempt to recover a language invariant political attribute of the models.

The remainder of the paper is organized as follows. Section 2 situates the paper in the literature on ideological evaluation, multilingual LLM behavior, measurement invariance, and political information intermediaries. Section 3 describes the experimental design and measurement strategy. Section 4 presents the main results. Section 5 examines sensitivity to questionnaire translation and wording, together with instrument dependence. Section 6 discusses the implications for multilingual evaluation and audit design, and Section 7 concludes.

## 2 Related Literature and Contribution

Research on the ideological evaluation of LLMs has largely relied on political questionnaires and survey batteries to place model responses in interpretable ideological spaces. Using the Political Compass, Motoki et al. (2024) document a position in the libertarian left for ChatGPT and extend their analysis to Brazil and the United Kingdom by asking the model to impersonate voters from those countries. Because all three exercises rely on prompts in English, the scope of these inferences depends on whether ideological measurement remains comparable across languages. Rozado (2024) applies a similar measurement approach to a broader set of models, while Feng et al. (2023) connect political behavior in downstream applications with biases present in pretraining data. Related work compares ideological placement across model generations, asking whether newer models appear systematically more or less aligned with particular political viewpoints (Rozado 2024). The GPT-3.5 estimates in Motoki et al. (2024) have also served as a reference for interpreting subsequent generations. Since much of this evidence comes from evaluations in English, differences attributed to model evolution implicitly assume that the language used for evaluation does not alter the ideological measurement.

A broader literature on multilingual LLM behavior documents performance differences across languages (Blasi et al. 2022), weaker safety alignment in languages with fewer resources (Shen et al. 2024), and social biases associated with the data used to train these models (Navigli et al. 2023). More recent studies examine political evaluation across languages directly. Helwe et al. (2025) administer the Political Compass Test to fifteen models across the fifty most populous countries and their 33 official languages, varying both the prompting language and an assigned nationality. Changes in language produce placements dispersed about two and a half times as far from the centroid of each model as changes in nationality, with smaller models displaying greater stability. Because the questionnaires are machine translations, however, differences in language cannot be separated from differences introduced by translation, and the analysis is based on aggregate placements rather than matched comparisons at the item level.

Using official translations of the Political Compass in fourteen languages, Gurgurov et al. (2025) evaluate seven open models and add eleven paraphrased prompt templates around each unchanged statement as a robustness exercise. Larger models tend to move toward the libertarian left, while the number of language pairs with significantly different placements increases with model size. A small set of languages, including Slovenian, Turkish, Polish, and Persian, either resist this movement or shift toward more centrist or authoritarian right positions. Nadeem et al. (2026) reuse the benchmark covering 50 countries from Helwe et al. (2025) to evaluate GPT-3.5, GPT-5, Claude, Gemma, Mistral-7B, and DeepSeek-7B. They report systematic differences in responses to identical statements across languages, with German responses more progressive and English responses slightly more conservative on the economic axis among the European languages considered, and subsequently propose an activation steering method that reduces variation across languages. Evidence from another instrument is provided by Rettenberger et al. (2025), who compare open models with German political parties using voting advice statements and find that smaller models are closer to neutral in English than in German.

Not all studies find substantial differences. Weeber et al. (2026) evaluate the six multilingual models, out of fifteen initially screened, that pass their robustness criteria using 239 voting advice statements in German, English, French, Spanish, and Italian. Aggregate differences across languages are negligible, with significant differences limited to a small number of policy issues, including greater support for environmental protection in German, Spanish, and Italian than in English. Political alignment based only on English data subsequently shifts responses almost uniformly across the five languages. The existing evidence therefore indicates that language can affect political measurement, while also showing that the magnitude and direction of the differences depend on the instrument, the languages considered, and the questionnaire version.

These findings connect directly with the literature on measurement invariance. In psychometrics, observed score differences are difficult to interpret as differences in an underlying construct unless the instrument measures that construct equivalently across groups or languages (Vandenberg and Lance 2000; Chen 2008). Comparative political measurement raises the same concern when respondents do not use a common scale in the same way, since observed differences across groups can combine variation in the underlying construct with differential item functioning (King et al. 2004). The analysis in this paper does not implement a full latent variable test of measurement invariance in the psychometric sense. Its focus is instead operational: whether widely used ideological measurements support comparable conclusions when either the language or the version of the instrument changes. Evidence that measured political orientation also varies with the ideological instrument itself (Barmettler 2026) further makes instrument choice part of the measurement problem.

A related perspective comes from the public choice and political economy literature on the supply of political information. Models of media markets distinguish slant associated with audience demand, as when readers prefer news consistent with their prior beliefs (Mullainathan and Shleifer 2005) or newspaper slant follows reader preferences while ownership accounts for little of the variation (Gentzkow and Shapiro 2010), from slant generated on the supply side through the incentives of content producers (Baron 2006) or the capture of media outlets by interested parties (Besley and Prat 2006). LLMs create a related environment in which one provider serves several linguistic constituencies through the same branded system while alignment, localization, and political risk management are costly. Differences in incentives across language markets are therefore one possible source of asymmetric political output. Demand-side factors may also play a role, and their effects may be difficult to distinguish from those of supply-side mechanisms. Differences in displacement across providers may arise from provider-specific decisions regarding alignment and localization, but they may also reflect differences in training data or in the composition of users served by each provider within a given language. Therefore, we consider both channels as potential explanations, without attempting to disentangle their respective contributions.

Evidence from Weeber et al. (2026) provides a counterpoint to the supply interpretation. Preference alignment based only on English data shifts political responses almost uniformly across five Western languages, a pattern that would reduce rather than generate asymmetries across languages. Their setting, however, differs from the present analysis in both instrument and design, using statements from voting advice applications, models aligned through direct preference optimization, and five closely related European languages. Uniform transfer from one alignment intervention does not imply that the residual displacement under the full alignment regime used by each provider must also be common across providers. The heterogeneity documented in Section 4 describes this residual pattern, without indicating which channel generates it. Since the analysis observes neither the allocation of alignment effort within firms nor differences in training data and user composition across providers and languages, we consider both supply- and demand-side factors as potential explanations rather than an identified causal channel. The evidence concerns the comparability of political measurements and the descriptive patterns discussed in Section 6.

Against this literature, the contribution is primarily methodological and evaluative rather than a claim of first discovery or greater scale. Matched comparisons at the item level and genuinely stochastic repetitions allow sensitivity to questionnaire composition to be distinguished from variability across repeated runs. Independent Spanish and French translations, alternative wording in all three languages, and a second ideological instrument then provide separate variation in dimensions that multilingual audits often combine. This design makes it possible to assess how much of the measured difference across languages survives changes in translation, wording, and instrument, and how large those differences are relative to the variability of model responses. It also shows whether the resulting changes are large enough to alter comparisons across models and whether the sign and magnitude of the measured displacement depend on the ideological instrument. The paper neither proposes a new ideological measure nor attempts to recover the "true" politics of a model; its contribution is to establish which comparative conclusions remain stable as the measurement environment changes.

## 3 Experimental Design and Measurement

This section presents the experimental setup, the ideological instrument, the response format and decoding parameters, and the measurement strategy used to assess whether measured ideological placement varies with the language of elicitation.

### 3.1 Models, languages, and evaluation conditions

The analysis covers five large language models under a common protocol: GPT-3.5, GPT-4o, and GPT-5.2 from OpenAI, together with Mistral Large and Qwen-2.5. The three OpenAI generations correspond to different training periods and alignment regimes, which allows comparisons across generations within the same model family as well as language comparisons within each model. Mistral Large and Qwen-2.5 extend the analysis beyond a single model family and country of origin. Since model behavior may change over time even when the public model name remains unchanged, the exact model identifier returned by the API is recorded for every call. Table S3 in the Electronic Supplementary Material (ESM) reports the model snapshots, providers, collection dates, and decoding settings, while ESM Section S3.1 reproduces the system prompts and user message templates verbatim.

Each model is evaluated in English (EN), Spanish (ES), and French (FR) under the same experimental conditions. The instrument, number of repetitions, item order, and decoding parameters are identical across the three languages, so the Spanish versus English, French versus English, and Spanish versus French comparisons follow the same design. All language comparisons are made within models while the remaining experimental dimensions are held constant. Inputs are processed directly in the language in which they are submitted, without an explicit translation step in the pipeline. Because internal representations and the composition of training data are not observed, the analysis does not attribute the resulting differences to a particular mechanism, such as differences in training data, attention patterns associated with particular languages, or alignment procedures after training.

In the main specification, models receive only the questionnaire instructions and are not assigned a political identity. We refer to this as the *neutral* condition. Politically labeled identities are linked to particular national political contexts, and the same label may evoke different associations across languages. Assigning one could therefore combine the effect of the language of elicitation with differences in the political meaning activated by that identity. For this reason, the neutral condition is used as the benchmark for evaluating language invariance. Evaluations that assign an explicit political identity address a different question, namely whether political framing interacts with language, and are therefore excluded from the main estimates. ESM Section S8 examines this separately by administering the questionnaire under five Argentine political identities in all three languages and under the same protocol, and comparing the resulting shifts with the neutral benchmark.

### 3.2 Ideological instrument

Ideological positioning is measured using the full Political Compass questionnaire, which consists of 62 statements. Following standard practice, 18 items measure the economic axis, covering attitudes toward markets, redistribution, and the role of the state, while 43 items measure the social axis, covering attitudes toward authority, individual freedom, and social norms.[1]

For each language, the analysis uses the official version of the Political Compass. The 62 items are administered in the same fixed order across languages, runs, and models. Item order cannot affect the language comparisons because every statement is submitted through an independent API call containing only the system instructions and that statement, with no conversation history. Responses to previous items therefore cannot influence subsequent ones through priming, attempts to maintain consistency, or fatigue. Keeping the order identical across languages and runs also prevents differences in item sequence from introducing an additional source of variation.

Official versions of the same political questionnaire in different languages need not be psychometrically identical. The main design therefore identifies differences in measured ideological placement associated with the language and questionnaire version jointly; by itself, it cannot determine whether a difference comes from model behavior, from the translation of the instrument, or from an interaction between the two. Section 5 examines this issue by administering independent machine translations of the English original into Spanish and French and paraphrased versions in all three languages. These exercises test whether the measured differences persist under alternative versions of the questionnaire rather than attempting to decompose them into separate language and translation components.

### 3.3 Response format, decoding, and scaling

Each model completes the 62 item questionnaire in each of the three languages. For every statement, the model must select one of four response categories: *Strongly Disagree* (SD), *Disagree* (D), *Agree* (A), or *Strongly Agree* (SA), and return the answer as a single integer. These categories are mapped deterministically into the ordinal scale

$$x_{mlrq} \in \{0,1,2,3\},$$

where $m$ indexes models, $l$ languages, $r$ repetition runs, and $q$ questionnaire items. The values 0, 1, 2, and 3 correspond, respectively, to $SD$, $D$, $A$, and $SA$. Only responses that can be unambiguously parsed into $\{0,1,2,3\}$ are retained. Outputs that do not satisfy this requirement are treated as missing and excluded, as described in Section 3.4.

Responses are sampled at temperature 0.5, with completions limited to five tokens, which is sufficient for the single integer requested by the prompt. All other sampling parameters remain at the default values of each provider, and every combination of model and language is evaluated in four independent runs. The positive temperature is important because deterministic decoding would produce essentially the same response when the same prompt is administered repeatedly, leaving no meaningful information about response variability. Treating such repetitions as independent observations would therefore overstate the effective sample size. With a positive temperature, each run instead provides a stochastic draw from the response distribution, making it possible to measure variation in ideological placement across repeated administrations. The resulting variation across runs provides a benchmark for evaluating the magnitude of the differences across languages. The corresponding standard deviations are reported in Appendix A.

[1] One item carries zero weight on both dimensions and is not classified.

Raw ordinal responses cannot be compared directly across items because Political Compass statements differ in their directional loadings and intensity. Following Motoki et al. (2024), each response category is assigned weights that determine its contribution to the economic and social axes (ESM Section S1, Table S1).

For each item $q$, let $w_q^E = (w_{q0}^E, w_{q1}^E, w_{q2}^E, w_{q3}^E)$ and $w_q^S = (w_{q0}^S, w_{q1}^S, w_{q2}^S, w_{q3}^S)$ denote the economic ($E$) and social ($S$) weight vectors. Given a realized response $x_{mlrq}$, the corresponding contributions are

$$e_{mlrq} = w_{q,x_{mlrq}}^E \qquad and \qquad s_{mlrq} = w_{q,x_{mlrq}}^S.$$

Let $Q_E$ and $Q_S$ denote the sets of items with nonzero weights on the economic and social axes, respectively, with $|Q_E| = 18$ and $|Q_S| = 43$. To obtain the position of each model in a given language, the contribution of each item is first averaged over the $R = 4$ runs,

$$\bar{e}_{mlq} = \frac{1}{R}\sum_{r=1}^{R} e_{mlrq},$$

after which the contributions are summed within each axis:

$$E_{ml}^{sum} = \sum_{q\in Q_E} \bar{e}_{mlq}, \qquad S_{ml}^{sum} = \sum_{q\in Q_S} \bar{s}_{mlq}.$$

The resulting sums are converted into Political Compass coordinates using the affine transformations reported by Motoki et al. (2024):

$$E_{ml} = \frac{E_{ml}^{sum}}{8} + 0.38, \qquad S_{ml} = \frac{S_{ml}^{sum}}{19.5} + 2.41.$$

These transformations place the weighted sums on the standard Political Compass coordinate scale used in prior work. They are not estimated from the data in this paper and are applied directly to preserve comparability with the scoring procedure in Motoki et al. (2024).

### 3.4 Experimental structure and data construction

Across the five models, three languages, four runs, and 62 questionnaire items, the data collection plan called for 3,720 API calls. Each item was handled through a separate two message exchange, and every call generated a record containing the response, the exact model snapshot, the UTC timestamp, the decoding parameters, and token usage. A total of 3,712 calls were completed. Eight calls, all involving Mistral Large, continued to return rate limit errors despite repeated retries. One additional observation, obtained from a French item, produced a nonnumeric response that could not be mapped into $\{0,1,2,3\}$. No value is imputed for any of these cases. Excluding them leaves 3,711 usable responses, equivalent to 99.8 percent of the intended collection. The distribution of these cases across models and languages is reported in ESM Section S2 (Table S2).

The comparisons between languages are formed only after matching observations by model and questionnaire item. An item contributes to a particular language comparison when valid responses are available for that model in both languages involved. As a result, every comparison refers to the same item within the same model. Missing observations may reduce the number of items available for a given comparison, but they cannot by themselves create a difference between languages.

### 3.5 Measuring language dependence

Differences in ideological placement across languages are calculated within the same model and questionnaire items. For model $m$, axis $a \in \{E, S\}$, and an ordered pair of languages $(l, l')$, the difference is defined as

$$\Delta^a_{m,ll'} = \frac{1}{c_a} \sum_{q \in Q_a^{m,ll'}} \left( \bar{w}^a_{mlq} - \bar{w}^a_{ml'q} \right), \qquad (1)$$

where $\bar{w}^a_{mlq}$ is the contribution of item $q$ to axis $a$ after averaging over the four runs, as defined in Section 3.3. The set $Q_a^{m,ll'}$ contains the items on axis $a$ with valid responses in both languages, while $c_E = 8$ and $c_S = 19.5$ are the scale factors used for the economic and social dimensions. The affine constants in the Political Compass transformation disappear when differences are taken, so $\Delta^{a}_{m,ll'}$ is already expressed in Political Compass units. For every model and axis, the analysis considers ES − EN, FR − EN, and ES − FR.

Sensitivity to the composition of the questionnaire is assessed by resampling the matched items within each axis with replacement. Equation (1) is recalculated in each of $B$ = 10,000 replications while preserving the pairing between languages. Since the Political Compass is a fixed set of items rather than a random sample from a larger population of questions, the resulting intervals are not interpreted as conventional measures of sampling uncertainty. Instead, they indicate how dependent an estimated difference is on the particular items included in the questionnaire and whether the result is broadly distributed across the axis or concentrated in a small number of statements. One, two, and three stars identify estimates for which the 90, 95, and 99 percent percentile intervals, respectively, exclude zero.

Variation generated by the stochastic repetitions is evaluated separately. The difference across languages is recalculated for each of the four runs, and the standard deviation of these four estimates is used to summarize variation across repeated administrations of the same questionnaire. This provides a second benchmark, distinct from sensitivity to questionnaire composition, for judging the size of the estimated differences across languages relative to the variation observed when the experiment is repeated. Both measures are reported for the main analysis and the translation exercises in Appendix A.

## 4 Results

### 4.1 Ideological positioning by model and language

We begin by examining how measured ideological placement varies across languages at the level of the aggregate Political Compass coordinates. Table 1 reports the economic and social coordinates of the five models under elicitation in English, Spanish, and French, while Figure 1 displays the resulting fifteen positions in the two-dimensional ideological space.

Table 1: Ideological positioning by model and language of elicitation

| | English | | Spanish | | French | |
|---|---|---|---|---|---|---|
| Model | Econ. | Social | Econ. | Social | Econ. | Social |
| GPT-3.5 | −5.651 | −3.872 | −3.526 | −3.231 | −0.870 | 0.307 |
| GPT-4o | −5.683 | −5.282 | −5.745 | −6.744 | −6.620 | −7.693 |
| GPT-5.2 | −5.714 | −6.090 | −5.808 | −6.667 | −7.433 | −6.923 |
| Qwen-2.5 | −6.276 | −6.026 | −6.276 | −6.436 | −5.964 | −6.654 |
| Mistral Large | −4.870 | −5.346 | −6.589 | −6.552 | −6.183 | −7.385 |

*Notes:* This table reports Political Compass coordinates for each model–language pair under neutral prompting. Scores are constructed as described in Section 3.3 and averaged over four independent runs per model–language pair. Lower

values indicate more left-wing positions on the economic axis and more libertarian positions on the social axis. Evaluations use the official questionnaire version in each language.

Figure 1: Political Compass positioning by model and language of elicitation

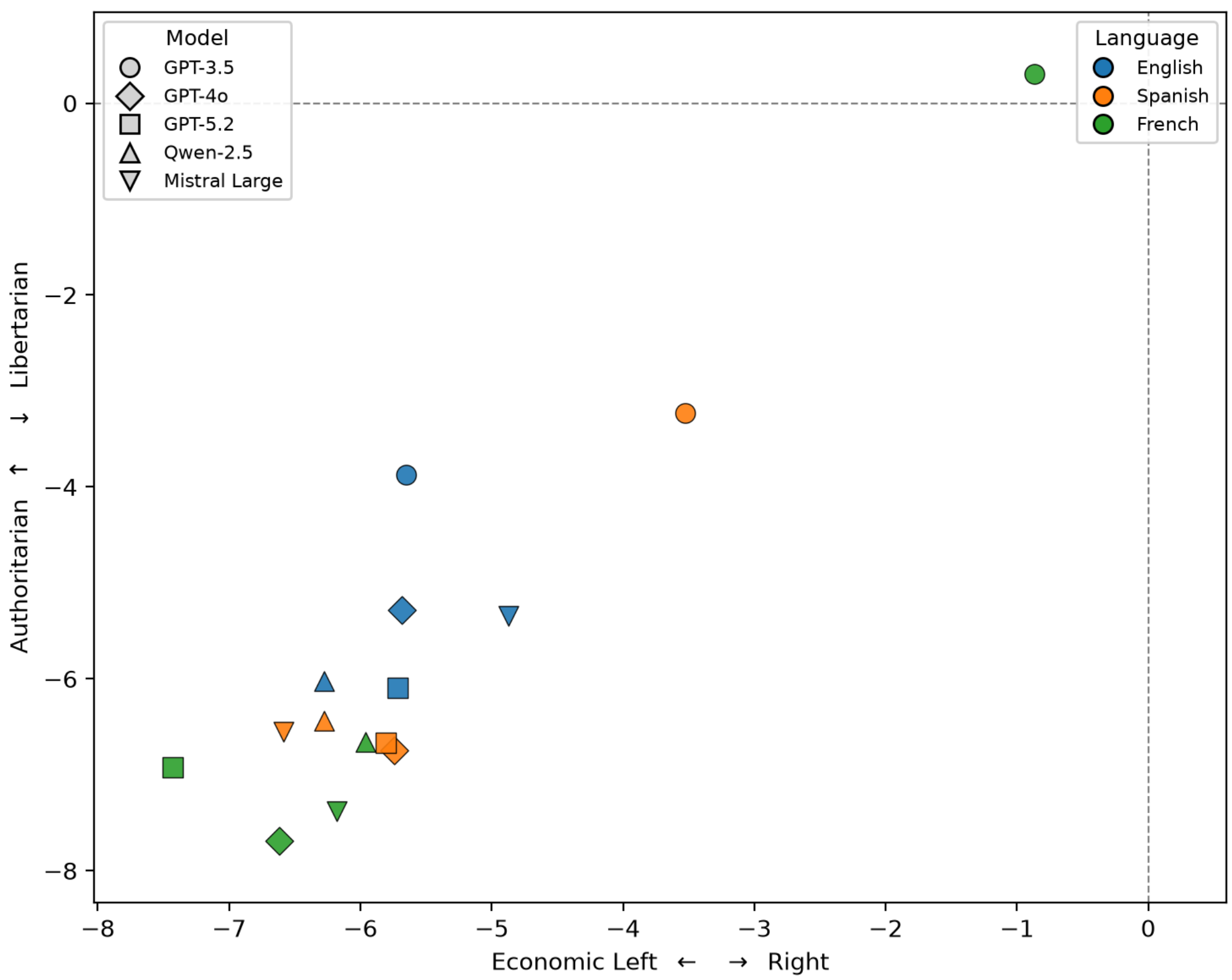

*Notes:* This figure plots the Political Compass coordinates reported in Table 1 for each model–language pair under neutral prompting. Coordinates are averaged over four independent runs. Marker shapes identify models, and colors indicate the language of elicitation. Lower economic values correspond to more left-wing positions, while lower social values correspond to more libertarian positions. Dashed lines indicate the origin.

The broad ideological classification is relatively stable across languages, but the exact position of a model is not. All five models remain on the left side of the economic axis in English, Spanish, and French, and fourteen of the fifteen model–language placements fall on the libertarian side of the social axis. The extent of the movement across languages, however, differs substantially across models. GPT-3.5 exhibits the largest change in the aggregate coordinates: its economic score moves from −5.651 in English to −0.870 in French, while its social score moves from −3.872 to 0.307. The French evaluation of GPT-3.5 is therefore the only one of the fifteen placements that falls outside the libertarian side of the social axis. At the other extreme, Qwen-2.5 remains much more tightly clustered, with economic coordinates ranging from −6.276 to −5.964 and social coordinates from −6.654 to −6.026. The remaining three models, GPT-4o, GPT-5.2, and Mistral

Large remain in the left-libertarian quadrant in all three languages, although their coordinates also shift with the language of elicitation.

Table 9 in Appendix A reports the dispersion of these positions across the four repetition runs for each model and language, together with the within-language correlation between the economic and social scores. The correlation is higher in Spanish and French than in English.

### 4.2 Language-induced shifts: matched-pair estimates

The coordinates reported in the previous subsection already show that the measured position of a given model can vary across languages. We now express those differences directly as within-model shifts and assess their sensitivity to questionnaire composition. Using the matched comparisons defined in Equation (1), Table 2 reports the difference in Political Compass coordinates for every model, ideological axis, and language pair (ES − EN, FR − EN, and ES − FR).

Table 2: Language-induced ideological shifts between all language pairs

| Model | Axis | ES − EN | FR − EN | ES − FR |
|---|---|---|---|---|
| GPT-3.5 | Economic | $2.125^{**}$ | $4.781^{***}$ | $-2.656^{***}$ |
| GPT-3.5 | Social | 0.641 | $4.179^{***}$ | $-3.538^{***}$ |
| GPT-4o | Economic | −0.062 | −0.938 | 0.875 |
| GPT-4o | Social | $-1.462^{**}$ | $-2.410^{***}$ | 0.949 |
| GPT-5.2 | Economic | −0.094 | $-1.719^{***}$ | $1.625^{*}$ |
| GPT-5.2 | Social | −0.577 | $-0.833^{*}$ | 0.256 |
| Qwen-2.5 | Economic | 0.000 | 0.312 | −0.312 |
| Qwen-2.5 | Social | −0.410 | −0.628 | 0.218 |
| Mistral Large | Economic | $-1.719^{***}$ | −1.312 | −0.406 |
| Mistral Large | Social | $-1.205^{***}$ | $-2.038^{***}$ | 0.833 |

*Notes:* This table reports estimated cross-language shifts in Political Compass coordinates under neutral prompting; see Section 3.5 for details. Positive values indicate a more right-leaning position on the economic axis and a more authoritarian position on the social axis in the first language relative to the second. $^{*}$, $^{**}$, and $^{***}$ indicate that the 90, 95, and 99 percent paired item-resampling intervals, respectively, exclude zero. These intervals capture sensitivity to questionnaire composition rather than conventional sampling uncertainty. Full intervals and run-level variability are reported in Table 7.

The estimates vary substantially across models, axes, and language pairs. The largest shifts occur for GPT-3.5 in the French–English comparison, where the economic coordinate is 4.781 Political Compass units higher in French than in English and the social coordinate is 4.179 units higher. Other models move in the opposite direction. For GPT-5.2, for example, the French–English shift is −1.719 on the economic axis, while for Mistral Large it is −2.038 on the social axis. Qwen-2.5 displays much smaller differences in the main specification, with no shift whose item-resampling interval excludes zero. The Spanish–English comparison shows the same absence of a common direction: GPT-3.5 shifts toward higher economic values, Mistral Large toward lower ones, while several estimates for the remaining models are close to zero. Cross-language differences therefore do not take the form of a common displacement shared by all models.

To assess the sensitivity of these estimates to questionnaire composition, we use item resampling. Several of the shifts reported in Table 2 remain on the same side of zero under item resampling, indicating that they are not driven by a small subset of questions. Because the Political Compass is a fixed battery rather than a random sample of items, these intervals measure robustness to questionnaire composition rather than conventional sampling uncertainty. The full intervals are reported in Table 7 in Appendix A.

We then compare the magnitude of the estimated shifts with the stochastic variation observed across repeated runs. Across the four runs, the largest standard deviation of any cross-language shift is 1.09 Political Compass units, and every shift whose 95 or 99 percent item-resampling interval excludes zero is more than twice as large as its own standard deviation across runs (Table 7 in Appendix A). The shifts that are most robust to questionnaire composition are therefore also large relative to their run-to-run variation.

These cross-language differences are also large enough to affect comparisons across models. Relative to the English ranking, between 40 and 50 percent of pairwise model rankings are reversed in Spanish or French, depending on the language and axis (see Appendix B). Reversals are less frequent between the two non-English languages: when comparing Spanish and French directly, four of the ten pairs reverse on the economic axis and only one on the social axis. These rankings are based on estimated mean positions across four runs, and some models are separated by very little: GPT-4o and Mistral Large differ by 0.064 units on the English social axis. The ordering of models placed this closely need not be stable across runs, so a reversal between them reflects an indeterminate relative position rather than a robust change. Thus, the relative ordering of models can change with the language in which the evaluation is conducted.

The Spanish–English shifts are also concentrated in a subset of questionnaire items. The five largest contributors account for 57 percent of the total absolute contribution on the economic axis and 38 percent on the social axis, and some of them differ substantively between the English and official Spanish versions. To check whether such translation differences drive the overall pattern, we separately audited all 62 items for changes in meaning, independently of their contribution to the shift and before any response data were examined. The ten items flagged by this audit — only partly overlapping with the high-contributing items just discussed — account for 17 percent of the total absolute contribution on the economic axis and 22 percent on the social axis, and excluding them leaves unchanged the sign of every Spanish–English shift whose item-resampling interval excludes zero. The full item-level decomposition is reported in ESM Section S4, with the analysis of the highest-contributing items in ESM Section S6.

We further examine whether the cross-language pattern changes when models are given an explicit political identity rather than evaluated under neutral prompting. For this exercise, we use five political personas from the Argentine context (peronista, radical, libertario, macrista, and kirchnerista) and administer the questionnaire under each persona in English, Spanish, and French. Cross-language non-invariance remains present, but the persona-conditioned shifts do not move in parallel with the neutral estimates. Qwen-2.5 has no shift whose interval excludes zero under neutral prompting, yet such shifts appear under two personas for the Spanish–English comparison and under four personas when all three language pairs are considered. Conversely, the robust neutral shifts for Mistral Large have no corresponding persona-conditioned estimates. Political framing therefore changes the pattern of cross-language displacement rather than introducing a common offset. The full persona results are reported in ESM Section S8.

## 5 Language versus Questionnaire Version

The results in Section 4 establish that measured ideological placement varies with the language of elicitation when the official version of the instrument is used in each language. Those estimates alone, however, cannot determine whether the observed differences arise because model responses change with the prompting language or because the official translations differ from the English questionnaire in ways that alter the content of individual items. A nonzero Spanish–English shift is consistent with either source of variation.

To examine this distinction, we vary the questionnaire version while holding the relevant language comparison fixed. Section 5.1 replaces the official Spanish and French questionnaires with alternative translations, allowing the sensitivity of the cross-language shifts to the particular translation to be assessed. Section 5.2 instead changes the wording within English, Spanish, and French while holding the language fixed and aiming to preserve substantive meaning. Section 5.3 then moves beyond alternative versions of the same questionnaire by changing the ideological instrument itself.

### 5.1 The translation arm

We generated independent Spanish and French translations of the English questionnaire using DeepL.[2] Each version was translated directly from the unchanged English original. The alternative translations differ substantially from the official questionnaire versions. For example, in Spanish, 55 of the 62 items are rendered differently.

[2] DeepL is a neural machine translation system.

Both translated questionnaires were administered under the same experimental protocol as the main analysis, in separate language-specific sessions. This allows us to examine whether the cross-language patterns documented above persist under alternative translations of the same instrument. Collection dates, model snapshots, and decoding settings are reported in ESM Section S3.

For each model and axis, we recompute the three language comparisons of Table 2 using the machine-translated Spanish and French questionnaires in place of the official versions. Because the alternative translations were administered in separate sessions, these comparisons are informative about sensitivity to translation choice but do not separately identify language and translation effects.

Table 3: Shifts under the independent machine translations

| Model | Axis | Machine ES − English | Machine FR − English | Machine ES − Machine FR |
|---|---|---|---|---|
| GPT-3.5 | Economic | 2.656** | 5.625*** | −2.969*** |
| GPT-3.5 | Social | 0.333 | 3.487*** | −3.154*** |
| GPT-4o | Economic | −0.625 | −1.917*** | 1.292* |
| GPT-4o | Social | −1.372*** | −2.000** | 0.628 |
| GPT-5.2 | Economic | 0.250 | −1.625*** | 1.875*** |
| GPT-5.2 | Social | −0.244 | −0.987** | 0.744* |
| Qwen-2.5 | Economic | 0.594 | −0.469 | 1.062*** |
| Qwen-2.5 | Social | −0.321 | −1.000* | 0.679 |
| Mistral Large | Economic | −1.188 | −1.750** | 0.562 |
| Mistral Large | Social | −1.321*** | −2.269*** | 0.949 |

*Notes:* This table reports cross-language shifts in Political Compass coordinates using independent machine translations of the English questionnaire into Spanish (Machine ES) and French (Machine FR) under neutral prompting. English refers to the original English questionnaire used in the main analysis. The machine-translated questionnaires were administered in separate sessions; see Section 5.1 for details. Stars are defined as in Table 2. Full intervals and run-level variability are reported in Table 8.

The cross-language pattern is largely reproduced under the alternative translations. The shifts obtained with the machine-translated questionnaires are highly correlated with those based on the official versions: the correlations are 0.94 and 0.96 for Spanish–English, above 0.98 for French–English, and 0.95 and 0.98 for Spanish–French on the economic and social axes, respectively (Table 3). Of the thirteen shifts whose item-resampling intervals exclude zero in Table 2, twelve retain their sign and remain robust under the alternative translations. The exception is the economic Spanish–English shift for Mistral Large, which falls from −1.719 to −1.188 and is no longer robust to item resampling.

Translation choice nevertheless matters for some individual estimates. Six shifts that are not robust under the official questionnaire become robust under the alternative translations. These include the economic French–English shifts for GPT-4o and Mistral Large. In the economic Spanish–French comparison for Qwen-2.5, the shift also changes sign, from −0.312 under the official version to 1.062 under the alternative translation.

Table 4: Response-level agreement between questionnaire versions: translation arm

| Model | $ES_{off}$ vs $ES_{MT}$ | $FR_{off}$ vs $FR_{MT}$ |
|---|---|---|
| GPT-3.5 | 0.927 | 0.829 |
| GPT-4o | 0.948 | 0.947 |
| GPT-5.2 | 0.955 | 0.968 |
| Qwen-2.5 | 0.948 | 0.944 |
| Mistral Large | 0.934 | 0.929 |
| *All models (pooled)* | 0.946 | 0.944 |

*Notes:* This table reports Pearson correlations across the 62 matched items between mean item-level responses under the official and machine-translated versions of each language. Responses are averaged over four runs. The pooled estimates combine all model–item observations. See Section 5.1 for details.

The correspondence is also strong at the item level. Mean responses under the official and machine-translated versions correlate between 0.93 and 0.97 for nearly every model and language combination, with GPT-3.5 in French as the main exception (0.83; Table 4). The broad cross-language pattern is therefore not specific to the official questionnaire versions, although individual model–axis estimates remain sensitive to translation choice. Full item-resampling intervals and run-level variation are reported in Table 8 in Appendix A.

### 5.2 The paraphrase arm: wording versus language

The translation exercise varies the rendering of the instrument across languages. A separate concern is that some of the measured cross-language differences may reflect sensitivity to the particular phrasing of individual items rather than to language itself. If substantively equivalent statements elicit different responses when they are worded differently within the same language, then part of the variation attributed to language may instead arise from the wording of the questionnaire. We therefore reword the statements in English, Spanish, and French while aiming to preserve their substantive content.

We generated the paraphrased questionnaires by round-trip machine translation through German using DeepL. Because German is outside the set of study languages, it provides a common and mechanical way to vary wording while limiting researcher discretion. The resulting versions differ from their official counterparts in 53 of 62 English items, 57 of 62 Spanish items, and 59 of 62 French items. The round trip leaves nine English, five Spanish, and three French items unchanged, with a median lexical overlap with the official version of 0.63, 0.57, and 0.68, respectively; the unchanged English items include 28 and 43, two of the highest-contributing social items identified in ESM Section S6.

Table 5 reports the cross-language shifts obtained with the paraphrased questionnaires.

Table 5: Cross-language shifts under the paraphrased questionnaire

| Model | Axis | $ES_{par} - EN_{par}$ | $FR_{par} - EN_{par}$ | $ES_{par} - FR_{par}$ |
|---|---|---|---|---|
| GPT-3.5 | Economic | −0.031 | 2.594** | −2.625*** |
| GPT-3.5 | Social | −0.731 | 3.821*** | −4.551*** |
| GPT-4o | Economic | −1.906** | −2.031 | 0.125 |
| GPT-4o | Social | −1.410* | −2.423*** | 1.013 |
| GPT-5.2 | Economic | −1.031* | −3.062*** | 2.031*** |
| GPT-5.2 | Social | −0.192 | −0.192 | 0.000 |
| Qwen-2.5 | Economic | −1.469*** | −0.500 | −0.969 |
| Qwen-2.5 | Social | −1.256*** | −0.308 | −0.949* |
| Mistral Large | Economic | −1.562** | −1.531 | −0.031 |
| Mistral Large | Social | −1.231** | −1.667** | 0.436 |

*Notes:* This table reports cross-language shifts in Political Compass coordinates using the paraphrased questionnaire versions under neutral prompting. Positive values indicate a more right-leaning position on the economic axis and a more authoritarian position on the social axis in the first language relative to the second. Stars are defined as in Table 2. See Section 5.2 for details.

Several of the differences observed under the official wording persist, particularly in the French–English comparison. For GPT-3.5, the economic and social shifts remain positive and robust to item resampling, at +2.594 and +3.821, respectively, compared with +4.781 and +4.179 under the official wording. The negative social shifts for GPT-4o and Mistral Large also persist, as does the negative economic shift for GPT-5.2.

The Spanish–English comparison is more sensitive to wording. The positive economic shift for GPT-3.5 under the official questionnaire (+2.125) disappears under paraphrase (−0.031). Rewording does not simply attenuate the estimated differences, however. Under the paraphrased questionnaire, the economic Spanish–English shift is negative and robust to item resampling for four of the five models, whereas under the official wording this pattern appears only for Mistral Large. The cross-language pattern therefore persists, but individual model–axis estimates can change substantially with the wording of the questionnaire.

Table 6 provides a complementary comparison at the item level.

Table 6: Response-level agreement between questionnaire versions: paraphrase arm

| Model | EN vs $EN_{par}$ | ES vs $ES_{par}$ | FR vs $FR_{par}$ |
|---|---|---|---|
| GPT-3.5 | 0.813 | 0.820 | 0.790 |
| GPT-4o | 0.935 | 0.896 | 0.919 |
| GPT-5.2 | 0.947 | 0.973 | 0.958 |
| Qwen-2.5 | 0.954 | 0.946 | 0.956 |
| Mistral Large | 0.914 | 0.952 | 0.955 |
| *All models (pooled)* | 0.920 | 0.927 | 0.944 |

*Notes:* This table reports Pearson correlations across the 62 matched items between mean item-level responses under the official and paraphrased versions of each language. Responses are averaged over four runs, and the two

questionnaire versions were administered in separate sessions. The pooled estimates combine all model–item observations. See Section 5.2 for details.

Responses to the official and paraphrased versions remain strongly related. The correlations range from 0.79 to 0.97 across model and language combinations, with pooled correlations of 0.920 in English, 0.927 in Spanish, and 0.944 in French.

Sensitivity to wording is also visible within languages. In English, the economic coordinate changes by between +0.906 and +2.469 for four models when the paraphrased questionnaire is used (ESM Section S7, Table S9). Because these comparisons use responses collected separately, they are not interpreted as pure wording effects.

Cross-language non-invariance therefore persists under alternative wording, but the magnitude and, in some cases, the direction of individual shifts remain sensitive to questionnaire phrasing. Measured ideological placement is thus not invariant either to the language of elicitation or to the particular wording of the instrument.

### 5.3 Beyond versions: the instrument itself

The translation and paraphrase exercises vary the questionnaire version while holding the underlying instrument fixed. A final question is whether the cross-language differences persist when the instrument itself changes. We therefore repeat the analysis using the IDRlabs Political Coordinates Test.

Cross-language non-invariance remains present, but the pattern differs from that obtained with the Political Compass. Under IDRlabs, the economic Spanish–English shift is leftward and robust to item resampling for GPT-5.2, Qwen-2.5, and Mistral Large. For GPT-3.5, by contrast, the rightward economic shift observed with the Political Compass is absent. Mistral Large is the only one of these three models for which the economic Spanish–English shift is leftward under both instruments. The full results are reported in ESM Section S5.

Changing the instrument therefore does not eliminate cross-language differences, but it changes which models exhibit them as well as their magnitude and, in some cases, their direction. What persists across instruments is the lack of cross-language invariance, rather than a stable model-specific displacement associated with a particular language.

## 6 Discussion

The central contribution of the paper is to treat ideological evaluation of LLMs as a problem of comparability across languages. Although the five models remain in broadly similar regions of the Political Compass when evaluated in English, Spanish, and French, the coordinates assigned to them vary with the language of elicitation. Some of these differences are large enough to change comparisons across models. These results should not be read as evidence that a model literally holds different political beliefs in different languages. Rather, the ideological position recovered by an audit depends in part on the language in which the measurement is conducted.

### 6.1 Language dependence is a measurement property

Matched comparisons of the same questionnaire items reveal no uniform displacement associated with a particular language. Both the size and the direction of the differences vary across models and ideological dimensions. The translation and paraphrase exercises provide further evidence on how this variation should be interpreted. Similar differences appear when independently produced Spanish and French translations replace the official questionnaire versions, which rules out an explanation based solely on those official translations. At the same time, some estimates for particular combinations of model and axis change when the wording or translation changes.

Taken together, these results imply that language cannot be treated simply as a neutral vehicle for measuring a fixed ideological position. They also do not isolate a causal effect of language from all other features of the questionnaire administered in that language. The measured outcome reflects both the language of elicitation and the particular version of the instrument. Accordingly, the evidence concerns dependence of the measurement exercise on language and questionnaire version rather than a language effect that is independent of the instrument.

This interpretation does not require models to possess stable political preferences analogous to those of human respondents. The relevant finding is that the conversion of model responses into an ideological score is sensitive to the language and version of the questionnaire. In that sense, the audit protocol itself helps determine the ideological position that the instrument records.

### 6.2 What changes and what does not

The results distinguish between stability in broad ideological classification and instability in relative placement. Across languages, the measured coordinates remain, on average, in the left and libertarian region of the Political Compass. What varies more substantially is the position of one model relative to another. Changing the language can narrow or enlarge the distance between models and can also change their ordering along an ideological axis. Relative to English, between four and five of the ten pairwise model rankings are reversed in Spanish or French. However, some of these reversals occur between models with very similar estimated positions, suggesting that their relative ordering may vary across runs. Comparisons among model families or generations based on a single evaluation language may therefore depend materially on that choice even when the models remain in the same broad ideological region.

Results from the IDRlabs Political Coordinates Test reinforce this distinction. Differences across languages remain present when ideological placement is measured with the alternative instrument, but neither their magnitude nor their direction is reproduced systematically from the Political Compass. What survives the change in instrument is therefore the lack of comparability across languages, rather than a consistent direction in which a given language moves the measured ideological position.

### 6.3 Implications for evaluation and deployment

Ideological evaluations conducted in several languages should preserve the language dimension rather than collapse it into a single score. Reporting separate placements by language avoids treating results obtained in one language, particularly English, as representative of the others. Comparisons are also more informative when the displacement between languages is shown alongside the average ideological coordinates and the questionnaire version is identified explicitly. Claims about "the ideology" of a model should therefore be qualified when they rest on evidence from only one language and one instrument.

The pattern of displacement is also relevant to how multilingual differences may arise in deployment. Section 2 discusses two potential mechanisms. On the demand side, responses in Spanish or French may reflect the political content of training data and the characteristics of users associated with each language. On the supply side, these differences may arise from providers' decisions about how to allocate alignment and localization efforts across language markets. Since training corpora and user bases also differ across providers, both mechanisms may explain why shifts in estimated ideological positions vary across providers. The descriptive evidence, however, does not allow us to distinguish between these explanations.

The results show that shifts in estimated ideological positions differ across models within a given language. In the Spanish versus English comparison reported in Table 2, GPT-3.5 shifts to the economic right (+2.125), while Mistral Large shifts to the economic left (−1.719). GPT-4o exhibits a shift only on the social axis (−1.462), whereas Qwen-2.5 shows no shift whose item-resampling interval excludes zero. These differences are also observed across successive versions from the same provider. For French versus English, the economic displacement of OpenAI models changes from +4.781 for GPT-3.5 to −0.938 for GPT-4o and −1.719 for GPT-5.2. Similarly, social displacement changes from +4.179 to −2.410 and −0.833, respectively. The disappearance of the large displacement of GPT-3.5 is also observed with the independent French translation (Table 3), although the relative economic magnitudes of GPT-4o and GPT-5.2 differ across the two versions. These findings suggest that language-related differences in measured ideology are model-specific and may change substantially across successive versions of the same product.

Outside the OpenAI family, two observations describe where the language differences are largest and smallest. Mistral Large displays the largest French displacement, −2.038 on the social axis, and its provider is headquartered in the jurisdiction of the EU Artificial Intelligence Act. Qwen-2.5 is the only model for which no shift has an interval that excludes zero, and English, Spanish, and French are all peripheral markets for its provider. Each observation can be read through either of the mechanisms discussed in Section 2: regulatory exposure and market focus could affect how alignment effort is allocated across languages, while differences across providers in the volume of text available for training in each language and in the composition of their user bases could produce the same pattern. With five models, and with neither channel observed, these observations do not discriminate between the two accounts; they indicate the kind of provider-level characteristics that either account would need to incorporate.

These results have two implications. First, evaluating a model only in English, or relying on evaluations of earlier versions, may provide limited information about the political content encountered by users in other languages. Second, identifying the mechanisms behind these differences would require additional information that is not available in the present analysis, such as how firms allocate alignment efforts across languages or how training data and user composition vary across providers and language markets. Ensuring comparability across languages is therefore an important first step toward understanding whether and why the political content provided by language models differs across linguistic groups.

### 6.4 Limitations and next steps

Because the analysis is restricted to five models, three languages, and two ideological instruments, its conclusions should be read within the boundaries of that design and as descriptive evidence rather than causal estimates. Neutral prompting facilitates comparisons across languages by avoiding political identities whose meaning may vary across national and linguistic contexts, although it cannot represent the broader range of prompts through which users interact with LLMs. Evidence from ESM Section S8, where assigning political identities changes the measured displacement across languages, further indicates that the patterns obtained under neutral prompting need not extend unchanged to other forms of political framing.

Similar limits apply to the distinction between language and the particular form of the questionnaire. Results from the independent translations and paraphrased versions show that the main lack of comparability does not depend on a single questionnaire version, while the variation across those exercises also prevents language from being separated completely from wording and instrument design. The estimates should consequently not be interpreted as a pure causal effect of language. Their scope is also tied to the model snapshots and API settings documented in ESM Section S3, since subsequent updates may alter the behavior observed in the experiment.

Extending the analysis to additional languages and prompting contexts would broaden the range over which these conclusions can be evaluated, as would the use of further ideological batteries with validated scoring across languages and the combination of model output comparisons with evidence on training, alignment, or deployment. These extensions would refine the empirical scope without changing the methodological issue identified here: language forms part of the conditions under which ideological measurement is obtained and should not be assumed to transmit an otherwise unchanged ideological signal. Evaluations conducted only in English can remain informative, but they provide incomplete evidence about the behavior of multilingual models.

## 7 Conclusion

Ideological evaluations of large language models are not invariant to the language in which they are conducted. Across English, Spanish, and French, the same models receive different Political Compass coordinates, and those differences are large enough to alter comparisons between them. Relative to English, four or five of the ten pairwise model rankings are reversed in Spanish or French, depending on the language and axis considered, while several of the larger displacements also exceed the variation observed across repeated administrations of the questionnaire. Although the models generally remain within similar broad ideological regions, the comparative conclusions drawn from the audit can therefore depend materially on the language of evaluation.

Independent Spanish and French translations reproduce the broader failure of comparability, making differences between the official questionnaire versions an insufficient explanation for the main result. Rewording the questionnaire also leaves differences across languages in place, although the magnitude and, for some combinations of model and axis, the direction of the displacement changes. A second ideological instrument produces the same lack of invariance without reproducing the same pattern of shifts. Taken together, these exercises support a measurement interpretation in which ideological placement depends on the model, the language of elicitation, the questionnaire version, and the instrument used. They do not support treating the resulting score as a language independent political attribute of the model, nor do they isolate a pure causal effect of language from the other features of the measurement environment.

The consequences extend beyond the design of ideological benchmarks because multilingual LLMs increasingly operate as political information intermediaries for constituencies that use the same product through different languages. A user normally observes the outputs available in the language being used rather than the treatment received by other linguistic groups, while providers face costly decisions over alignment, localization, and political auditing across markets. Under those conditions, asymmetries across languages may be difficult for users to observe and need not attract the same degree of provider attention in every constituency. The heterogeneous displacements documented across providers and model generations are compatible with such differences in incentives, as well as with differences in the training data and user composition each provider faces, although the design does not identify whether the observed pattern originates on the demand or supply side. What can be established from the present evidence is that evaluating political provision across linguistic constituencies requires measurements that are comparable across languages; without them, differences relevant for provider accountability remain difficult to detect.

Ideological audits should accordingly preserve language as part of the measurement environment rather than pool it away or treat it as a neutral channel. Reporting placements separately by language, identifying the questionnaire version and ideological instrument, and showing differences within the same model across languages makes clear which conclusions survive changes in the evaluation setting. Evidence obtained only in English can still be informative, but it describes one part of the behavior of a multilingual system rather than providing a language independent characterization of "the ideology" of the model. The broader lesson is therefore not that a model acquires a different political identity when the language changes, but that claims about its ideological position are only as portable as the measurement procedure used to produce them.

## Appendix A: Item-Resampling Intervals and Run-Level Variability

This appendix reports the two measures underlying the markers of Tables 2 and 3, together with the run-level dispersion of the positions of Table 1. For every model, axis, and language pair, Table 7 reports the 95 percent paired item-level bootstrap percentile interval, which measures sensitivity to questionnaire composition, and the standard deviation of the shift across the four repeated runs, which measures behavioral variability, for the main trilingual session; Table 8 reports the same quantities for the three machine-translation shifts of Section 5.1. Both measures are defined in Section 3.5.

Table 7: Item-resampling intervals and run-level variability: language shifts of Table 2

| | | 95% item-resampling interval | | | SD across runs | | |
|---|---|---|---|---|---|---|---|
| Model | Axis | ES − EN | FR − EN | ES − FR | ES − EN | FR − EN | ES − FR |
| GPT-3.5 | Economic | [0.219, 4.469] | [1.938, 7.938] | [−4.594, −0.969] | 0.979 | 1.091 | 0.616 |
| GPT-3.5 | Social | [−0.872, 2.141] | [2.782, 5.654] | [−4.962, −2.179] | 0.279 | 0.570 | 0.433 |
| GPT-4o | Economic | [−1.375, 1.625] | [−3.688, 2.125] | [−2.000, 3.750] | 0.375 | 0.780 | 0.425 |
| GPT-4o | Social | [−2.795, −0.218] | [−4.154, −0.910] | [−0.308, 2.321] | 0.194 | 0.268 | 0.403 |
| GPT-5.2 | Economic | [−1.375, 1.469] | [−3.126, −0.500] | [0.000, 3.750] | 0.739 | 0.438 | 0.354 |
| GPT-5.2 | Social | [−1.487, 0.282] | [−1.808, 0.115] | [−0.590, 1.103] | 0.391 | 0.524 | 0.178 |
| Qwen-2.5 | Economic | [−0.688, 0.688] | [−0.844, 1.562] | [−1.438, 0.625] | 0.530 | 0.298 | 0.239 |
| Qwen-2.5 | Social | [−1.487, 0.679] | [−1.910, 0.654] | [−0.795, 1.218] | 0.399 | 0.391 | 0.440 |
| Mistral Large | Economic | [−2.938, −0.656] | [−3.188, 0.938] | [−2.594, 1.438] | 0.237 | 0.315 | 0.188 |
| Mistral Large | Social | [−1.987, −0.513] | [−3.538, −0.654] | [−0.462, 2.218] | 0.339 | 0.380 | 0.294 |

*Notes:* This table reports the 95 percent item-resampling intervals and run-level standard deviations associated with the language shifts in Table 2. The intervals are obtained from the paired item-level bootstrap with B = 10,000 replications and capture sensitivity to questionnaire composition rather than conventional sampling uncertainty. Run-level standard deviations summarize variation in the estimated shifts across the four repeated administrations. See Section 3.5 for details.

Table 8: Item-resampling intervals and run-level variability: machine-translation shifts

| | | 95% item-resampling interval | | | SD across runs | | |
|---|---|---|---|---|---|---|---|
| Model | Axis | ES − EN | FR − EN | ES − FR | ES − EN | FR − EN | ES − FR |
| GPT-3.5 | Economic | [0.312, 5.250] | [2.625, 8.969] | [−5.562, −0.844] | 0.483 | 0.540 | 0.766 |
| GPT-3.5 | Social | [−0.667, 1.321] | [2.231, 4.833] | [−4.628, −1.756] | 0.129 | 0.453 | 0.524 |
| GPT-4o | Economic | [−1.719, 0.438] | [−3.917, −0.438] | [−0.188, 3.031] | 0.941 | 0.461 | 1.058 |
| GPT-4o | Social | [−2.500, −0.385] | [−3.705, −0.397] | [−0.667, 2.051] | 0.194 | 0.382 | 0.253 |
| GPT-5.2 | Economic | [−1.125, 2.188] | [−2.938, −0.594] | [0.281, 4.062] | 0.468 | 0.395 | 0.368 |
| GPT-5.2 | Social | [−1.269, 0.833] | [−1.949, −0.090] | [−0.051, 1.667] | 0.500 | 0.400 | 0.129 |
| Qwen-2.5 | Economic | [−0.469, 1.812] | [−1.562, 0.656] | [0.156, 2.375] | 0.524 | 0.844 | 0.657 |
| Qwen-2.5 | Social | [−1.487, 0.846] | [−2.064, 0.090] | [−0.295, 1.654] | 0.227 | 0.066 | 0.227 |
| Mistral Large | Economic | [−3.125, 1.062] | [−3.250, −0.188] | [−1.375, 2.594] | 0.239 | 0.213 | 0.216 |
| Mistral Large | Social | [−2.282, −0.436] | [−3.731, −0.846] | [−0.179, 2.256] | 0.387 | 0.590 | 0.273 |

*Notes:* This table reports the 95 percent item-resampling intervals and run-level standard deviations associated with the machine-translation shifts in Table 3. ES and FR refer to the independently machine-translated Spanish and French questionnaires, while EN refers to the English questionnaire from the main session. Intervals and run-level standard deviations are constructed as in Table 7; see Section 3.5 for details.

Table 9 complements these shift-level measures with the dispersion of the positions themselves. For each model and language it reports the variance of the Political Compass score across the four repetition runs, separately by axis; the bottom row reports the Pearson correlation between economic and social scores pooled across all model–run combinations within each language. Because the neutral condition is the primary specification and decoding is stochastic at temperature 0.5, cross-run variance is the natural measure of within-model behavioral dispersion, and it is the quantity against which the cross-language shifts of Section 4 are compared.

Table 9: Cross-run variance and axis correlation by language

| | Economic axis | | | Social axis | | |
|---|---|---|---|---|---|---|
| Model | EN | ES | FR | EN | ES | FR |
| GPT-3.5 | 0.077 | 0.493 | 0.760 | 0.234 | 0.160 | 0.014 |
| GPT-4o | 0.141 | 0.000 | 0.062 | 0.030 | 0.089 | 0.023 |
| GPT-5.2 | 0.171 | 0.141 | 0.016 | 0.151 | 0.032 | 0.035 |
| Qwen-2.5 | 0.483 | 0.066 | 0.181 | 0.067 | 0.120 | 0.020 |
| Mistral Large | 0.062 | 0.004 | 0.016 | 0.144 | 0.029 | 0.000 |
| $r(E,S)$ by language: EN $= 0.227$; ES $= 0.872$; FR $= 0.930$ | | | | | | |

*Notes:* This table reports the variance of Political Compass coordinates across the four repeated runs for each model, language, and axis. The bottom row reports the Pearson correlation between economic and social coordinates within each language, pooling model–run observations.

## Appendix B: Magnitude of the Language Effect

This appendix evaluates the substantive magnitude of the cross-language differences using three complementary measures: pairwise ranking inversions, standardized shifts, and ratios relative to observed ideological dispersion. These measures capture different aspects of the size of the language effect and avoid relying on a single benchmark that may be sensitive to how closely the models are clustered on a given axis.

### B.1 Pairwise ranking inversions

With five models, each axis yields ten pairwise comparisons. For each pair, we ask whether the model with the higher score in English remains higher in the comparison language. A pair is classified as inverted when this ordering reverses. Rankings are based on the estimated mean positions reported in Table 1, averaged across four runs. Because some models have very similar mean scores, their relative ordering may vary across runs. Thus, reversals between closely positioned models may reflect uncertainty in their relative rankings rather than robust differences.

Table 10 reports the number of inversions for the official and machine-translated Spanish and French questionnaires, using the English positions in Table 1 as the reference. Relative to English, between 40 and 50 percent of pairwise rankings reverse. When comparing Spanish and French directly, the official questionnaires yield fewer reversals: four of the ten pairs on the economic axis and one pair on the social axis (GPT-5.2 and Mistral Large). The pattern also persists across questionnaire versions: three economic and three social inversions occur under both Spanish versions, while three economic and four social inversions occur under both French versions. In Spanish, the economic inversions that persist across versions all involve Mistral Large, while GPT-4o appears repeatedly among the persistent social-axis inversions in both Spanish and French.

Table 10: Pairwise model-ranking inversions across languages

| Comparison | Economic | Social |
|---|---|---|
| English → Spanish (official) | 4/10 | 4/10 |
| English → French (official) | 4/10 | 5/10 |
| English → Spanish (machine) | 5/10 | 5/10 |
| English → French (machine) | 4/10 | 4/10 |
| Robust to both Spanish versions | 3 | 3 |
| Robust to both French versions | 3 | 4 |

*Notes:* This table reports pairwise model-ranking inversions across languages for each ideological axis. With five models, there are ten pairwise comparisons per axis. An inversion occurs when the relative ordering of two models is reversed between English and the comparison language. Official rankings are based on Table 1, while machine-translation rankings use the corresponding positions from Section 5.1. The robust rows report inversions that occur under both the official and machine-translated versions of the same language. Rankings are based on estimated mean positions averaged across four runs. For models with very similar scores, the relative ordering may vary across runs.

### B.2 Standardized effects

Table 11 expresses each Spanish–English shift relative to two alternative dispersion measures. The first standardizes the shift by the cross-model standard deviation of the English positions on the corresponding axis, providing a benchmark based on differences across models. The second uses the standard deviation of the Spanish–English shift across repeated runs, pooled across models, and therefore compares the estimated displacement with the stochastic variation observed across repeated evaluations.

Relative to run-to-run variability, the four Spanish–English shifts whose item-resampling intervals exclude zero in Table 2 are approximately 2.7 to 4.4 times as large as the corresponding noise benchmark. The cross-language differences that are most robust to questionnaire composition are therefore also large relative to the stochastic variation observed across repeated runs.

Table 11: Standardized Spanish–English shifts

| | Relative to cross-model dispersion | | Relative to run-to-run noise | |
|---|---|---|---|---|
| Model | Economic | Social | Economic | Social |
| GPT-3.5 | +4.24 | +0.72 | +3.38 | +1.94 |
| GPT-4o | −0.12 | −1.64 | −0.10 | −4.44 |
| GPT-5.2 | −0.19 | −0.65 | −0.15 | −1.75 |
| Qwen-2.5 | +0.00 | −0.46 | +0.00 | −1.24 |
| Mistral Large | −3.43 | −1.35 | −2.73 | −3.66 |
| Denominator | 0.501 | 0.893 | 0.630 | 0.330 |

*Notes:* This table reports the Spanish–English shifts from Table 2 standardized by two alternative benchmarks. Cross-model dispersion is the standard deviation of English-language positions across the five models, while run-to-run variability summarizes the dispersion of Spanish–English shifts across repeated runs. The denominators used for each axis are reported in the bottom row. See Section 3.5 and Table 7 for details.

### B.3 Ratio measures

Table 12 reports two additional scale-free measures of magnitude. $\text{Ratio}_{\text{range}}$ compares the mean absolute cross-language shift with the full range of ideological scores observed across models, languages, and runs. By this measure, changing the questionnaire language moves a model by roughly one tenth of the total ideological range observed in the study.

$\text{Ratio}_{\text{inter}}$ instead compares the same mean absolute shift with the average pairwise distance between models in English. This ratio is larger on the economic axis because the English economic positions are tightly clustered, with a mean pairwise distance of 0.575. The measure should therefore be interpreted together with its denominator rather than as a standalone indication of effect size.

Table 12: Ratio measures of substantive magnitude

| | Economic | Social |
|---|---|---|
| Mean absolute shift, ES − EN | 0.800 | 0.859 |
| $\text{Ratio}_{\text{range}}$ (primary) | 0.108 | 0.104 |
| denominator: observed score range | 7.375 | 8.256 |
| $\text{Ratio}_{\text{inter}}$ (secondary) | 1.391 | 0.829 |
| denominator: mean pairwise EN distance | 0.575 | 1.036 |

*Notes:* This table reports two measures of the substantive magnitude of the Spanish–English shifts in Table 2. Both use the mean absolute shift across models as the numerator. $\text{Ratio}_{\text{range}}$ standardizes this quantity by the observed range of Political Compass scores, while $\text{Ratio}_{\text{inter}}$ standardizes it by the average pairwise distance between English-language model positions. The corresponding denominators are reported in the table.

## Statements and Declarations

**Funding.** No funding was received to assist with the preparation of this manuscript.

**Competing interests.** The authors have no relevant financial or non-financial interests to disclose.

**Data availability.** The data generated and analyzed during the current study, together with the replication code necessary to reproduce the results, will be made available in a public repository upon publication.

**Code availability.** The replication code necessary to reproduce the tables and figures will be made available in a public repository upon publication.

**Use of AI-assisted tools.** Large language models are the object of analysis in this study. The model responses analyzed in the paper were generated through the experimental prompts and procedures described in the manuscript. AI-assisted tools were also used during manuscript preparation to support code checking, debugging, and language editing. All analytical decisions, interpretations, results, and final text were reviewed and approved by the authors, who take full responsibility for the content of the manuscript.